\documentclass[aps,preprint,nofootinbib]{revtex4}
\usepackage{graphicx}
\begin{document}
\title{\Large \bf Uniqueness of Rotating Charged Black Holes in Five-Dimensional Minimal Gauged Supergravity}
\author{\large Haji Ahmedov and Alikram N. Aliev}
\address{Feza G\"ursey Institute, \c Cengelk\" oy, 34684   Istanbul, Turkey}
\date{\today}

\begin{abstract}

We study a five-dimensional spacetime admitting, in the presence of torsion, a non-degenerate  conformal Killing-Yano $2$-form which is closed with respect to both the usual exterior differentiation  and the exterior differentiation  with torsion. Furthermore, assuming that the torsion is closed and co-closed with respect to the exterior differentiation  with torsion, we prove that such a spacetime is the only spacetime given by the Chong-Cveti\u{c}-L\"{u}-Pope  solution for stationary, rotating charged black holes with two independent angular momenta in five-dimensional minimal gauged supergravity.

\vspace{10mm}

\begin{center}
{\it Dedicated to Nihat Berker on the occasion of his 60th birthday}
\end{center}

\end{abstract}


\maketitle

\section{Introduction}

Uniqueness is one of the most striking  features of ``the truth and beauty" of black holes in all spacetime dimensions. In four dimensions, general relativity describes the final equilibrium state of black holes in terms of  stationary asymptotically flat exact solutions of spherical topology to the Einstein field equations. The fundamental property of these solutions is their uniqueness: In  most general case, a stationary and asymptotically flat black hole is uniquely characterized by the mass, angular momentum and the electric charge \cite{carter1, hawking1, robin} (see also \cite{heusler} and references therein). The uniqueness  has been a crucial  basis for studying  many remarkable properties of black holes, thereby constituting  a firm ground for their search in  the real universe. However, it turns out that the uniqueness property  is fundamental to black holes only in four dimensions and  it does not survive in higher dimensions.

For static and asymptotically flat vacuum black holes, the uniqueness of the Schwarzschild  solution can still be extended to higher spacetime dimensions \cite{gibida}, but it is not the case for rotating black holes.  For instance, in five dimensions  there exist rotating black hole solutions  with different horizon topologies: The Myers-Perry solution with spherical horizon topology  \cite{mp} and the Emparan-Reall black ring solution \cite{er} with the horizon topology of  $  S^2 \times S^1 $. These solutions  may have the same mass and angular momenta. Clearly, this fact breaches the black hole uniqueness in five dimensions. The lack  of black hole uniqueness  is also supported  by   a recent generalization of Hawking's theorem \cite{hawking1} to higher dimensions \cite{wald}. This generalization guarantees the existence of a higher-dimensional stationary black hole with a single rotational Killing symmetry, unlike the Myers-Perry solution, which possesses multi-rotational Killing symmetries. Thus, in general the higher-dimensional black holes  are not uniquely characterized  by their physical parameters, such as the mass and angular momenta. However, to classify these black holes one can still look for the uniqueness of each black hole solution separately. In particular, a uniqueness result along this line was achieved for a Myers-Perry black hole in five dimensions. Namely, it was proved that in five  dimensions, the only stationary, asymptotically flat black hole solution with two rotational symmetries and spherical topology of the horizon is given by the Myers-Perry metric \cite{mori, holls}.

It is  also a remarkable fact that the uniqueness results for  stationary black holes in four and higher dimensions are intimately related to the hidden symmetries of these black holes.  As is known,  stationary black holes in four dimensions  admit  a closed conformal Killing-Yano (CCKY) $2$-form which encodes all hidden symmetries generated by $2$-rank Killing-Yano and Killing tensors of these spacetimes \cite{wp,p,fkrev}. Using this fact, it was shown that the most general solution of the Einstein field equations with a cosmological constant which  admits the CCKY $2$-form is given only by the Kerr-NUT-(A)dS metric \cite{dietz, taxi}.  Recently, it was demonstrated that the higher-dimensional generalization of the Kerr-NUT-(A)dS spacetime  constructed  in \cite{clp} also admits  a CCKY  $2$-form which generates the tower of hidden symmetries in higher dimensions \cite{fk1, fkrev}. With this  CCKY  $2$-form, the authors of  \cite{kfk} managed to prove that  the higher-dimensional Kerr-NUT-(A)dS metric of \cite{clp} is a unique solution (see also \cite{houri}).

The aim of this Letter is to prove the similar uniqueness result for rotating  charged black holes in five-dimensional minimal gauged supergravity. The general solution with two independent rotational symmetries that describes  these black holes was found by Chong, Cveti\u{c}, L\"{u} and Pope (CCLP) \cite{cclp}. In a recent paper \cite{acif}, it was shown that this solution can be put in a Kerr-Schild type framework with two independent scalar functions, that provides its simple derivation. The gyromagnetic ratios of these black holes  were studied in \cite{aliev1}. The CCLP metric also admits hidden symmetries generated by a $2$-rank Killing tensor. This results in a complete separability of variables for the Hamilton-Jacobi  and Klein-Gordon equations \cite{dkl, ad}. The separability properties of the equation of motion for a stationary string in this  metric were examined in \cite{ahaci1}.  However,  the CCLP  metric does not admit the usual Killing-Yano tensor and therefore the separation of variables for the Dirac equation  \cite{ad0}, unlike its uncharged counterpart \cite{oota, wu1, wu2, ahaci2}. On the other hand, the author of \cite{wu3} showed  that such a separability can be  achieved  by adding a counter-term into the usual Dirac equation. The hidden symmetries underlying the separability of variables in the modified Dirac equation are governed by the generalized (``non-vacuum") Killing-Yano equation \cite{wu3}. A nice geometrical interpretation of this result  was given in \cite{kky}. The authors introduced  a torsion  $3$-form, defining  it as  the Hodge  dual of the Maxwell $2$-form. They showed that the CCLP metric admits a CCKY  $2$-form in the presence of the torsion and the associated $3$-rank Killing-Yano tensor which ensures  the separability of variables in the modified Dirac equation.

We prove that the only spacetime admitting a closed (with  respect to  both the usual differential operator and the differential operator with torsion) conformal Killing-Yano $2$-form in the presence of torsion is given by the CCLP metric, provided that the torsion is closed and co-closed  with respect to differential operators with torsion. We note that in  the asymptotically flat case, the uniqueness of rotating charged black holes in minimal ungauged supergravity was proved in \cite{tomi} by extending the boundary value analysis of \cite{mori}.

\section{The metric and its hidden symmetries}

The five-dimensional  minimal  gauged supergravity is governed by the Lagrangian
\begin{eqnarray}
{\cal L} &=& (R+ \Lambda)\ast 1 -\frac{1}{2} \,\, {\ast} F\,\wedge F +\frac{1}{3\sqrt{3}}\,\, F\,\wedge F\,\wedge A\,\,,
\label{5sugralag}
\end{eqnarray}
which results in the following system of Einstein-Maxwell-Chern-Simons field equations
\begin{equation}
{R_{\mu}}^{\nu}= \frac{1}{2}\left(F_{\mu \lambda} F^{\nu
\lambda}-\frac{1}{6}\,{\delta_{\mu}}^{\nu}\,F_{\alpha \beta}
F^{\alpha \beta}\right) - \frac{1}{3}\, \Lambda\,{\delta_{\mu}}^{\nu}\,,
\label{einmax}
\end{equation}
\begin{equation}
 d F=0\,~~~~~~ d\ast F-\frac{1}{\sqrt{3}} \,F\wedge F=0\,.
\label{maxw}
\end{equation}
As we have mentioned  above, the general rotating  charged black hole solution subject to these equations was   constructed in \cite{cclp}. It is interesting that this solution can be  written in the most simple form \cite{lmp} (see also \cite{kky}) by using the  ``canonical" basis $1$-forms given by
\begin{eqnarray}\label{canbasis}
 e^1 &=& \sqrt{\frac{x-y}{4X}}\,dx\,, ~~~~~e^2 = \sqrt{\frac{y-x}{4Y}}\,dy\,, \nonumber \\[2mm]
e^{\bar{1}} &= &\sqrt{\frac{X}{x(y-x)}}( dt+yd\phi)\,,~~~~~
 e^{\bar{2}} = \sqrt{\frac{Y}{y(x-y)}}( dt+xd\phi)\,, \nonumber \\[2mm]
 e^0 &=& \frac{1}{\sqrt{xy}}\left[\mu dt +\mu (x+y) d\phi + xy d\psi -yA_q-xA_p \right],
 \end{eqnarray}
where
\begin{eqnarray}
A_q &=& \frac{q}{x-y}(dt+y d\phi )\,,~~~~~  A_p = \frac{p}{y-x}(dt+x d\phi)\,,
\label{split}
\end{eqnarray}
such that $ Q=q-p $ and the electromagnetic potential $1$-form have the form
\begin{equation}
    A= \sqrt{3}(A_q + A_p)\,,~~~~~ F=dA\,.
\end{equation}
The  functions $ X $ and $ Y $  are given by
\begin{eqnarray}\label{function1}
X&=& (\mu+q)^2 + a_1 x + a_3 x^2 + \frac{\Lambda}{12}\,x^3\,,  \\[2mm]
    Y&=& (\mu+p)^2 + a_2 y + a_3  y^2 + \frac{\Lambda}{12}\,y^3\,.
\label{function2}
\end{eqnarray}
Thus, we have the metric in the form \footnote{We adopt the positive-definite signature for convenience.}
\begin{equation}
g= \sum_{a=1}^2 (e^a \,e^a
    + e^{\bar{a}}\,e^{\bar{a}}) +e^0\,e^0\,.
\label{g}
\end{equation}
This metric involves four free parameters related to the mass, electric charge and two angular momenta of the black hole. We note that the parameter $ a_3 $ in (\ref{function1}) and (\ref{function2}) can be eliminated  using the translations in the directions of  $ x $ and $ y $.

The authors of \cite{kky} suggested  a modification of the conformal Killing-Yano equation, introducing a torsion into the spacetime. In particular, a `` closed" conformal $2$-rank Killing-Yano (CCKY) tensor in this spacetime obeys the equation
\begin{equation}\label{gccky}
\nabla^T_\mu h_{\nu\rho}= g_{\mu\nu}\xi_\rho -
    g_{\mu\rho}\xi_\nu\,,
 \end{equation}
which implies that
\begin{equation}
    d^T h=0\,,~~~~~\xi=-\frac{1}{4}\, \delta^T h\,.
\label{close}
\end{equation}
Here the covariant derivative operator with torsion acting on a vector field $ V $ is defined as follows
\begin{equation}
\nabla^T_\mu V_\nu = \nabla_\mu V_\nu -\frac{1}{2} T^\sigma_{ \
    \mu\nu} V_\sigma\,,
    \label{covder}
\end{equation}
where $ T $ is the torsion $3$-form and $ \nabla_\mu $ is the usual
covariant derivative operator. Moreover, we have the metricity condition $ \nabla^T_\mu g_{\nu\rho}=0 $. Similarly, for a $3$-form field $ \Psi $ in five dimensions, we have
\begin{equation}
d^T \Psi= d \Psi - ({\ast T})\wedge ({\ast \Psi})\,.
\label{dtorsion}
\end{equation}
We note that  $ \delta^T $ is the adjoint of the exterior derivative operator with torsion $ d^T $. Further details of the differential operations with torsion can be found in  \cite{kky}.

Next, defining  the torsion  by the Hodge dual of the Maxwell $2$-form $ F= d A $ through the relation
\begin{equation}
T=\frac{1}{\sqrt{3}}\ast F\,,
\label{tfs}
\end{equation}
and using  the Maxwell-Chern-Simons equations (\ref{maxw}) along with (\ref{dtorsion}) and the fact that  $ \delta^T T= \delta T $, it is easy to show that  the torsion is ``harmonic" with respect to $ d^T $ and $ \delta^T $ operations. That is, we have
\begin{equation}
d^T T=0\,,~~~~~\delta^T T=0\,.
\label{harmonic}
\end{equation}
Remarkably, the spacetime (\ref{g}) admits a  non-degenerate CCKY  tensor $ (d^T h=0) $ \cite{kky}, which is  given by
\begin{eqnarray}
h&=&\sqrt{-x}\, e^{1}\wedge e^{\bar{1}}+\sqrt{-y} \,e^{2}\wedge e^{\bar{2}}\,.
\label{gccky1}
\end{eqnarray}
It is straightforward to verify that this tensor is also $d$-closed, $ d h=0 $.  It is important to note that the Hodge dual of this tensor is a  $3$-rank Killing-Yano tensor that explains the separability of variables for the Dirac equation \cite{wu3} in the metric (\ref{g}). Moreover, this tensor also results in a  $2$-rank  Killing tensor of this metric \cite{dkl, ad, kky}.

\section{The Uniqueness}

In this  section, we  wish to prove the uniqueness of the general rotating charged black hole solution of  five-dimensional  minimal  gauged supergravity, constructed by Chong, Cveti\u{c}, L\"{u} and Pope in \cite{cclp}. Namely, we prove the following

{\bf Theorem:} {\it Suppose a five-dimensional spacetime  admits, in the presence of torsion, a  non-degenerate  conformal Killing-Yano (CKY) $2$-form $ h $ which is both  $ d^T $ and  $ \,d $-closed and the torsion is harmonic, satisfying the conditions $\, d^T T=0 \,$ and $\, \delta^T T=0\,. $ Then, this spacetime is the only spacetime given by the Chong-Cveti\u{c}-L\"{u}-Pope solution for stationary, rotating charged black holes with two independent angular momenta in five-dimensional minimal gauged supergravity.}

We will present the proof of this theorem in several steps: (i) We begin by noting that a $2$-rank  antisymmetric tensor  $ h_{\mu\nu} $ on a metric space defines the linear map
\begin{equation}
    H\cdot v^\mu \equiv {h^\mu}_\nu v^\nu
    \label{map}
\end{equation}
and the `` eigenfunctions" of this operator given by
\begin{equation}\label{eig}
    H\cdot e_a^{ \ \mu} = -x_a \,e_{{\bar{a}}}^{ \ \mu}\,, ~~~~~ H\cdot e_{\bar{a}}^{ \ \mu} = x_a\,
    e_{a}^{ \ \mu}\,,~~~~~ H\cdot e_0^{ \ \mu} = 0\,, ~~~~ a=1,2\,.
\end{equation}
form a Darboux basis \cite{kfk}. The CKY $2$-form $ h $ determines an orthonormal Darboux basis, in which  one can diagonalize the metric
 $ g $  and `` skew"-diagonalize the $2$-form $ h $. We have
\begin{equation}
\label{h}
g=\sum_{a=1}^2 ( e^a e^a+
    e^{\bar{a}} e^{\bar{a}})+ e^0e^0\,,~~~~~
h=\sum_{a=1}^2 x_a e^a\wedge e^{\bar{a}}\,.
\end{equation}
Clearly, there still exists a freedom with respect to $ SO(2) $ rotations in $(e^a, e^{\bar{a}})$ $2$-planes and we can use this freedom to choose the vector filed $ \xi $ in equation (\ref{gccky}) as follows
\begin{equation}\label{K}
\xi^\mu=\sum_{a=1}^2 \sqrt{Q_a} \,e_{\bar{a}}^{\ \mu} + \sqrt{S}\,e_0^{ \ \mu}\,,
\end{equation}
where $ Q_a $ and $ S $ are unknown scalar functions.
For further convenience, it is also useful to use  the dual Darboux basis $ e_A $  with $ A=a,\, \bar{a},\, 0 $. In this notation,
equations (\ref{eig}) reduce to the form
\begin{equation}\label{H}
 H\cdot e_A^{ \ \mu} = Z_A \,e_{\bar{A}}^{ \ \mu}\,,
\end{equation}
where the eigenvalues
$$ Z_a=-x_a\,,~~~~  Z_{\bar{a}}= x_a \,,~~~~ Z_0=0\,. $$
Using now the closedness conditions  for CKY $2$-form $ h $,
\begin{equation}\label{closed}
 d^T h=0 \,,~~~~~d h=0\,,
\end{equation}
we find that the torsion obeys the following algebraic equations
\begin{equation}
    T_{A[BC}h^{A}_{\ D]}=0\,,
\end{equation}
where square brackets denote antisymmetrization. These equations  are solved by
\begin{equation}\label{tt}
T=  T_1 \,e^0\wedge e^1\wedge e^{\bar{1}} +T_2 \,e^0\wedge e^2\wedge
    e^{\bar{2}}\,.
\end{equation}
Later, we shall also  present   the explicit expressions for the components  $ T_1 $ and  $ T_2 $\,.

(ii) Next, using equation (\ref{gccky}) along with  (\ref{map}), we arrive at the equation
\begin{equation}\label{eq1}
d^T H \cdot e_A^{\ \mu} = \xi_A e^B e_B^\mu - e_A \xi^B
    e_B^{\ \mu}\,.
\end{equation}
Combining this equation with (\ref{H}) and taking into account the orthogonality condition  $ (e_A,e_B)=\delta_{AB}\, $, we find  that
\begin{equation}
\label{eq2}
d Z_A= \xi_A \,e^{\bar{A}}-\xi_{\bar{A}}\,e^A\,.
\end{equation}
This equation  along with  (\ref{K})  determines the gradient of the eigenvalues $ x_a $. We have
\begin{equation}\label{eq0}
    dx_a=\sqrt{Q_a}\, e^a\,.
\end{equation}

(iii) We shall now show that the CKY $2$-form $ h $ under consideration is constant along the associated vector field $ \xi $. We note that
\begin{equation}\label{Lie}
   \pounds_\xi h=d (\iota_\xi h)+ \iota_\xi   dh\,,
\end{equation}
where $ \iota_\xi $ is the interior product operator. Since  $ h $ is $d$-closed as well,  the second term  on the right-hand side vanishes. Using (\ref{eq0}), we have
\begin{equation}
\iota_\xi h = -\sum_{a=1}^2
x_a\sqrt{Q_a}e^a=-\frac{1}{2}d \left(\,\sum_{a=1}^2 x^2_a\right),
\end{equation}
which shows the first term on the right-hand side of (\ref{Lie}) is of an exact differential. Thus, we obtain that
\begin{equation}
\pounds_\xi h=0\,.
\label{hxi}
\end{equation}

Let us now assume that  $ \xi $ is the Killing vector. Then applying the Lie derivative to  equation (\ref{H}), we  obtain
\begin{equation}\label{eq3}
\pounds_\xi  e_A^\mu = -P_A e_A^\mu\,,
\end{equation}
where
\begin{equation}
    P_a=P_{\bar{a}}=i_\xi d\log \sqrt{Q_a}, \ \ \  P_0=i_\xi d\log
    \sqrt{S}\,.
\end{equation}
In obtaining these expressions we have used equations (\ref{eq2}) and  (\ref{eq0}). We note that equation (\ref{eq3}) can also be written in the alternative form
\begin{equation}\label{eq4}
    \partial_A \xi^C +(\omega_{AB}^C-\omega^C_{BA})\xi^B=
    \delta^C_A P_A
\end{equation}
and  the connection $1$-forms
\begin{equation}
\omega_{\ A}^C=\omega^C_{BA}\,e^B
\end{equation}
are  defined by the equation
\begin{equation}
    d e_A = \omega_{\ A}^C\wedge e_C\,.
\end{equation}
Next, we define a symmetric operator $ H^2= H\cdot H $, for which we have
\begin{equation}\label{eig1}
  -H^2   e_A^{ \ \mu} = Z_a^2 \,e_{{A}}^{ \ \mu}\,, ~~~~~
  -H^2    e_{\bar A}^{ \ \mu} = Z_a^2 \,e_{{{\bar A}}}^{ \ \mu}\,.
\end{equation}
Taking the usual covariant derivative of this equation, it is easy to show that
\begin{equation}\label{cc}
    \sum_{B} (Z_B^2-Z_A^2)\,\omega^B_{ \ A} e_B= dH^2 \cdot e_A +d\log
    Z_A^2 e_A\,.
\end{equation}
Combining this equation with (\ref{eq1}), we find that the connection $1$-forms are given by
\begin{equation}\label{eq5}
\omega^B_{ \ A}=-\frac{1}{2}T^{B}_{\ A} +\frac{Z_A (\xi^{\bar{A}}e^B
-\xi^B e^{\bar{A}})+ Z_{\bar{B}} (\xi^A e^{\bar{B}}-\xi^{\bar{B}}e^A
)}{Z_B^2-Z_A^2}\,,~~~~ A\neq \bar{B}\,,
\end{equation}
where $T_{AB}=e^C T_{CAB}\, $.
Using  this expression in (\ref{eq4}) we see  that for $  b\neq
    a $
\begin{equation}
    \partial_{0} Q_b=0\,,  ~~~~~\partial_{\bar{a}} Q_b=0\,,~~~~~ \partial_{\bar{a}} S=0
\end{equation}
and
\begin{equation}
    \omega^a_{\ \bar{a}}=-\frac{\partial \sqrt{Q_a}}{\partial x_a}\,e^{\bar{a}}+ \frac{\sqrt{S}}{x_a}\,e^0 +\sum_{b\neq
    a}\frac{x_a\sqrt{Q_b}}{x_a^2-x_b^2}\,e^{\bar{b}}-\frac{1}{2}\,T^a_{ \ \bar{a}}\,\,.
\end{equation}

These results enable one to calculate explicitly  the corresponding Lie derivatives of all basis $1$-forms. We have
\begin{equation}
\pounds_{e_{\bar{a}}} e_{\bar{b}}=0\,, \ \ \ \ \pounds_{e_{0}}
    e_{\bar{b}}=0\,, \ \ \ \ \ \pounds_{e_{\bar{a}}} e_{0}=0\,.
\end{equation}
With this in mind and  for $ \xi \neq 0 $, it follows from  equation (\ref{eq3}) that  $ P_ A=0 $. This justifies the assumption made above that $ \xi $ is the Killing vector. That is,
\begin{equation}
\pounds_\xi g=0\,.
\end{equation}

(iv) Substituting the quantities (\ref{eq5}) in equation (\ref{eq4}), we obtain the following equations
\begin{equation}\label{Q}
    \frac{\partial \sqrt{Q_a}}{\partial x_b}=\frac{x_b
    \sqrt{Q_a}}{x_a^2-x_b^2}\,, \  \  \  a\neq b\,,
\end{equation}
\begin{equation}\label{S}
    \frac{\partial \sqrt{S}}{\partial x_a}+\frac{\sqrt{S}}{x_a}=
    T_a\,.
\end{equation}
From  equations in (\ref{Q}) we easily find that
\begin{equation}
Q_1=\frac{X_1(x_1)}{x_1^2-x_2^2}\,, \  \  \ Q_2=\frac{X_2(x_2)}{x_2^2-x_1^2}\,,
\label{q1q2}
\end{equation}
where $ X_a(x_a) $ is an  arbitrary function. In order to solve equation (\ref{S}) we need the components of the torsion tensor. From the condition  $\, d^T T=0 \,$ we obtain that
 \begin{equation}
T_1x_1+T_2x_2=0\,,
\end{equation}
and
\begin{equation}
    \frac{\partial T_a}{\partial x_b}+\frac{T_a}{x_b} =2\,\frac{x_b T_a- x_a T_b
    }{x_a^2-x_b^2}\,, \ \ \  a \neq  b\,.
\end{equation}
The  solution to these equations is given by
\begin{equation}
    T_1=\frac{2Q x_2}{(x_1^2-x_2^2)^2}\,,~~~~~  T_2=- \frac{2Q x_1}{(x_1^2-x_2^2)^2}\,,
    \label{t1t2}
\end{equation}
where  $Q $ is an arbitrary constant. It is easy to check that with this solution the condition $\, \delta^T T=0 \,$ is  fulfilled  as well.
Using (\ref{t1t2}) in equation (\ref{S}) we find its solution in the form
\begin{equation}
    \sqrt{S} = \frac{\mu}{x_1x_2}+\frac{1}{x_1^2-x_2^2}\left(
    p\,\frac{x_1}{x_2}-q\,\frac{x_2}{x_1}\right)\,,
\end{equation}
where $\mu$\,, $ p $ and $ q$  are constants parameters and  $q-p=Q $\,.

(v) In the vacuum case with zero torsion,  one can construct all Killing vectors admitted by the spacetime, using only the fact of the existence of  a closed  conformal Killing-Yano tensor in this spacetime \cite{kfk}. For instance, in five dimensions in addition to the {\it primary} Killing vector $ \xi $,  we have two other Killing vectors given by
\begin{equation}
    \varphi^A= K^A_{ \ B}\xi^B\,, \ \ \ \ \
    \chi^A=\frac{1}{8}\,\varepsilon^{ABCDE}h_{BC}h_{DE}\,,
\end{equation}
where the Killing tensor
\begin{equation}
    K_{AB}= h_{AC}h^{C}_{ \ B}-\frac{1}{2} \,\delta_{AB}h^2\,.
    \label{kilt}
\end{equation}
However in the presence of torsion only $ \chi $ appears to be  the Killing vector. Indeed, using the
identity
\begin{equation}
    \nabla_{(A} \chi_{B)}=\nabla^T_{(A}\chi_{B)}
 \end{equation}
and  (\ref{gccky}) we find that
\begin{equation}
    \nabla^T_{(A} \chi_{B)}= \frac{1}{4}\varepsilon_{(AB)CDE}\xi^C h^{DE}=0\,,
\end{equation}
where round brackets  stand for symmetrization. On the other hand, using  (\ref{hxi}) it is straightforward to show that
\begin{equation}
\nabla_{(A} \varphi_{B)}=-\xi^C\nabla_C K_{AB}\,.
\end{equation}
Equations (\ref{gccky}) and (\ref{kilt}) enable us to put this equation in the form
\begin{equation}
    \nabla_{(A} \varphi_{B)}=\frac{1}{2}\xi^C( T_{AC}^{\ \ \ D} K_{DB}+ T_{BC}^{\ \ \ D} K_{DA})\,.
\end{equation}
Thus, it follows that in the presence of torsion, the information encoded in $ h $ is not enough to construct the whole set of  Killing vectors. Therefore, to construct the third Killing vector  one needs to invoke  the torsion as well. We assume  that the putative third Killing vector  has the form
\begin{equation}
    \xi_A= \varphi_A+ f \chi_A\,,
\end{equation}
where $ f = f(x_1\,,x_2) $ is a scalar function. Then, from the associated Killing equations we find that
\begin{eqnarray}
  \frac{\partial f}{\partial x_1} + \frac{x_1}{x_2}\,T_1 &=& 0  \,, \\[2mm]
  \frac{\partial f}{\partial x_2} + \frac{x_2}{x_1}\,T_2 &=& 0\,.
\end{eqnarray}
Substituting in these equations the expressions in (\ref{t1t2}), we find the simple solution
\begin{equation}
f= \frac{Q}{x^2_1-x_2^2}\,\,.
\end{equation}
Thus, the desired Killing vector is given by
\begin{equation}\label{K1}
    \eta = x_2^2 \sqrt{Q_1}\,e_{\bar{1}}+x_1^2 \sqrt{Q_2}\,e_{\bar{2}}+
    \left[ \sqrt{S} (x_1^2+x_2^2) + \frac{Q x_1x_2}{x_1^2-x_2^2}\right] e_0\,.
 \end{equation}
We can now choose the the coordinate system $ (t,
\phi, \psi)$, such that
\begin{equation}
    \xi=\partial_t\,, \ \ \ \  \eta =\partial_\phi, \ \ \ \
    \chi=\partial_\psi\,.
\end{equation}
and using equations (\ref{K}) and (\ref{K1}) together with  $ \chi=  x_1 x_2\,e_0 $, we find that
\begin{equation}
    e^{\bar{1}}=\sqrt{Q_1} (dt + x_2^2 d\phi )\,, \ \ \ \ e^{\bar{2}}=\sqrt{Q_2} (dt +x_1^2
    d\phi)
\end{equation}

\begin{equation}
    e^0 =x_1x_2 d\psi + \sqrt{S}\,dt+\left[
    \sqrt{S}(x_1^2+x_2^2)+\frac{Q x_1x_2}{x_1^2-x_2^2}\right] d\phi
\end{equation}
With these basis $1$-forms and those given by (\ref{eq0}) and (\ref{q1q2}), the metric in (\ref{h})  satisfies the field equations (\ref{einmax}) and (\ref{maxw}) of five-dimensional minimal gauged supergravity, if one takes
\begin{equation}\label{eqq}
x=-x_1^2\,,~~~~ y=-x_2^2\,, ~~~~ X_1=-\frac{X}{x_1^2}\,\,,~~~~
X_2=-\frac{Y}{x_2^2}\,\,, ~~~~~\phi \rightarrow -\phi\,
\end{equation}
and
\begin{equation}
F=\sqrt{3} \ast T\,.
\end{equation}
That is, it becomes precisely the same as the CCLP metric (\ref{g}) with the canonical basis (\ref{canbasis}). This completes the proof of the theorem.

\end{document}